\documentstyle[12pt]{article}

\title{Pattern Formation and Functionality in Swarm Models}
\author{ Erik M. Rauch$^1$,   Mark M. Millonas$^2$  and Dante R. Chialvo$^3$ \\
$^1$\small Yale University, P.O. Box
202200 Yale Station, New Haven, CT 06520 \\ $^2$\small Theoretical Division and
CNLS, MS B258 Los Alamos National Laboratory, \\ \small Los Alamos, NM 87545 \\
$^3$\small Division of Neural Systems, Memory and Aging, University of Arizona,
Tucson AZ 85724
  }

\begin{document}

\maketitle

\begin{abstract}
We explore a simplified class of models we call swarms, which are
inspired by the collective behavior of social insects.   We perform
a mean-field stability analysis and perform numerical simulations of the
model. Several interesting types of behavior emerge in the vicinity of a
second-order phase transition in the model, including the formation of stable
lines of traffic flow, and memory reconstitution and  bootstrapping.
In addition to providing an understanding of certain classes of biological
behavior, these models bear a  generic resemblance to a number of pattern
formation processes in the physical sciences.
\end{abstract}

\section{The Model}

The self-organization of patterns of flow in social insect swarms is a
beautiful example of how intelligent and efficient behavior of the whole
can be achieved even in the absence of any particular intelligence or
forethought
of the individuals.\cite{Wilson}  Indeed, such patterns can have functionality
even without the awareness of the individual entities themselves.  A
study of the essential elements of swarm dynamics provides an understanding of
such behaviors, and is our principal goal in this letter.  A secondary, broader
goal  is to study the generic behavior of a kind of stochastic particle-field
system in which the motion of the
``particles" both changes the field and is affected by the field. There are a
number of systems like this in the physical sciences.  We particularly have in
mind systems where the relaxation time scales of the field are slow in
comparison to the motion of the particles and this is what makes the present
work unique and interesting.

Consider the following scenario.  Suppose particles move in a
field $\sigma({\bf x})$  that field corresponds to an energy
$U[\sigma({\bf x})] = U({\bf x})$ so that each particle experiences a force
due to the field of ${\bf F} = -\nabla U({\bf x})$.  Suppose also that the
particles are subject to  random perturbations so that their motion is
only statistically in the direction of ${\bf F}$.  We could describe their
motion by the stochastic (Langevin) type equation
\begin{equation}
 \ddot{\bf x}+  \Gamma \dot {\bf x} +\nabla U[\sigma({\bf x})] = {\bf f}(t).
\end{equation}
where ${\bf f}(t)$ is Gaussian white noise with
$<{\bf f}(t)>= {\bf 0}$ and
$<{\bf f}(t) \cdot{\bf f}(s)>= (1/\beta){\bf
1}\delta(t-s)$. The parameter   $\Gamma$ is a friction coefficient. Obviously
for social insects this term is not to be taken literally.  The ``friction"  in
such  cases merely parameterizes the tendency of the organism-particles to
continue in a given direction.  A smaller ``friction" thus means that the
organism's velocity vector is correlated over longer times.  In physical
systems
this parameter may have a more literal meaning.

The noise
strength is $1/\beta$.  Thus $\beta$, which we call the osmotropotaxic
sensitivity\footnote{Osmotropotaxis is pheromone gradient following, as
explained
in greater detail in \cite{hang,o2,o3}}, determines the degree of determinacy
with which the organism-particle follow the gradient of the field
${\bf F}({\bf x}) =  -\nabla U({\bf x}) = -U^\prime[\sigma({\bf x})] \nabla
\sigma({\bf x})$.  If
$\beta$ is large the predominant force felt by the particles is ${\bf F}$, and
they follow ${\bf F}$ more accurately.  If $\beta$ is small the noise becomes
more important and the organism-particles follow the field gradient with less
certainty.  It is necessary to assume probabilistic laws in order to
reproduce the behaviors observed experimentally, and the osmotropotaxic
sensitivity can be, and has been, observed  experimentally.\cite{d1,d2,d3}  As
we
will see, the degree of randomness plays a very important role in the
functionality of the swarm.  Most probabably it is hardwired into the sensory
apparatus of the organism.

Up to this point we have simply described the noisy motion of particles in
some  ``energy landscape".  The principal element of interest here is that
the energy landscape will be allowed to evolve in response to the noisy motion
of the particles.  We will choose one of the simplest possible models for how
this occurs.\cite{m1,m2,m3}  The presence of a particle at ${\bf x}$ will
``deposit" an amount of field $\eta$ per unit time.  For insect swarms
$\sigma(\bf x)$ is to be interpreted as the pheromonal field density\cite{ww2}
at ${\bf x}$. An amount of field
$\kappa
\sigma({\bf x})$ will ``decay" per unit time.  The field might also diffuse in
space with diffusion constant $D$ such that the field will evolve according to
\begin{equation}
{\partial \sigma({\bf x})\over \partial t} = D\nabla^2\sigma({\bf x})
+\eta \rho({\bf x}) - \kappa \sigma({\bf x})
\end{equation}
where $\rho({\bf x})$ is the particle density at ${\bf x}$.  When the particles
are social insects this equation describes the evolution of
pheromonal field $\sigma$ laid down by the organisms as they walk, and  the
parameter $\kappa$ is an evaporation rate.

The only thing which has been left unspecified here is $U[\sigma]$.  In this
letter we use
\begin{equation}
U[\sigma] = -\ln\left(1 + {\sigma\over 1+\delta \sigma}\right).
\end{equation}
The use of this ``field energy function" has already been justified
elsewhere,\cite{m1,m2,m3} where it has been explained at length how {\it this
function is derived from the experimentally observed behavior of real ants}.
Other energy functions can be used for different physical situations, and the
general framework of this model is capable of supporting many different
types of systems.

 Here we call $1/\delta$ the capacity.  This models the fact that
 an actual ant's response to additional concentration of pheromone decreases
somewhat at high concentrations.  This gives rise to a peaked function for the
average time an ant will stay on a line of ant traffic as the concentration of
pheromone is varied -- a fact which has been repeatedly observed
experimentally\cite{p1,p2,p3}. This is
perhaps significant in light of one of the results presented here: trails and
networks do not spontaneously form in the absence of this saturation effect.
One
reasonable physiological explanation for this effect is that the ant's  sensing
organs become saturated: since each antenna has only a finite number of
pheromone
receptor sites, the antennae are effectively jammed at high concentration, and
the response to gradients can be expected to grow less
pronounced.

The model described above was constructed precisely because it
approximately reproduces the ``microscopic" behavior of individual ants as
observed in the laboratory.\cite{o1,o2,o3,d3} Notice that we have resisted the
temptation to write a Fokker-Planck equation for $\rho({\bf p},{\bf x},t)$ from
Eq. 1.  This is to emphasize that we do not want to consider the continous
limit,
but rather the case where internal fluctuations are significant.  Since the
field
$\sigma$ corresponds to the pheromone which, unlike the ant density $\rho$, is
composed of a macroscopically large number of particles, we can use the
continuous Equation 2 to describe it. We
have expressed the model in this particularly suggestive form to emphasize  its
physical aspects, and its generic relationship to other physical systems.  We
will not expand on this point here, but will merely indicate the generic
resemblence of this model to problems such as anomalous ionic diffusion in
polymeric materials, stochastic growth processes, the evolution of river
basins,
and other types of complex physical systems that incorporate mobile elements
and
an evolving  substrate of some kind.\cite{other}

\section{Analysis}

Since a detailed mathematical analysis of this type of system has already been
made\cite{m1,m2},  we will confine ourselves here to a brief statement of the
pertinent results.

We analyze the case where the relaxation of the field $\sigma({\bf x})$ is very
slow in comparison to the relaxation time of the particle density $\rho({\bf
x})$.  In this case the field is said to slave\cite{hakken} to the particle
denisty, and the particle density can be removed from the picture in the
following way.  Since the field changes very little on these time scales we
will
assume it is constant as the  particles  equilibrate in the energy landscape
$U[\sigma({\bf x})]$.  It is easy to show that the equilibrium distribution of
particles evolves to
\begin{equation}
\rho_e[\sigma({\bf x})] = {\exp(-\beta U[\sigma({\bf x})])\over\int\ d^D{\bf
y}\exp(-\beta U[\sigma({\bf y})]) }
\end{equation}
Elimination of the particle variables is made possible by substituting this
expression into Eq. 2, so that
\begin{equation}
{\partial \sigma({\bf x},t)\over \partial t} = {\eta N\exp(-\beta U[\sigma({\bf
x},t)])\over\int\ d^D{\bf y}\exp(-\beta U[\sigma({\bf y},t)]) } - \kappa
\sigma({\bf x},t),
\end{equation}
where $N$ is the total number of particles, and where we neglect diffusion and
will not consider it futher in the present letter. For our present purpose we
will only be interested in the stability of the uniform phase of the system,
$\sigma({\bf x},t)=\sigma(t)$, which we will treat by mean-field theory.   The
above equation has a mean-field solution of
$\sigma_0 = N\eta/V\kappa$, where $V = \int\ d^D{\bf y}$.

The stability of this solution is found by expanding about the solution with
$\sigma({\bf x}, t) =
\sigma_0 + \delta\sigma({\bf x},t)$. This leads to the mean-field stability
criterion
\begin{equation}
{\beta\sigma_0} U^\prime[\sigma_0] +1 >0.
\end{equation}
This is the most important theoretical result, and it gives the location of the
second-order phase transition, that is, the location of the boundary separating
totally random behavior from ordered behavior of varying types in the
physiological phase space.  As shown in \cite{m1,m2}, ordered behavior sets in
when this criterion is broken.  This criterion is true for any energy function
$U[\sigma]$, and allows us to calculate the physiological phase boundaries in
every case.

 For the particular behavioral energy function used
here (Eq. 3) the transition lies along the curve
\begin{equation}
\beta_c(\delta,\sigma_0) =   {1+ 1/\sigma_0+ 2\delta  +\delta \sigma_0 +
\delta^2\sigma_0}.
\end{equation}
The symmetry which allows for a mean-field type solution for the location of
the transition line case (detailed balance) is maintained up to the point of
the
transition and is effectively spontaneously broken at that point.  This means
the points of transition from disorder to order can be determined
theoretically,
 but not the resulting patterns, which must be
determined via simulations or further theoretical analysis.

In addition to being the major landmark in the physiological parameter space,
we believe both on general grounds, and because of the results presented in the
next sections, that the location of this line has significant behavioral
implications.  Clearly it is important that ordered behavior exists so that
ordered patterns of flow can form, but just as
importantly, the behavior should not be too rigid and ordered since fluctuation
and instabilities might increase the flexibility of response of the mass
action.
Thus, if the dynamics are such that there can be significant fluctuations in
the
patterns of mass action, the swarms could better respond to a changing
environment.  We might conclude that a ``good" place to be would be in the
order
region, but near to the transition line in such a way as to optimize the
conflicting tendencies of controlled order behaviors versus flexible random
behavior.  We will also see in the next sections that large fluctuation
actually
serves to stabilize some of the important patterns of collective behavior of
the
system.

\section{Numerics}

Equations 1, 2 and describe the system, but
for the purposes of simulation it is necessary to
introduce some discretization of space and time, and to translate the noisy
behavioral function observed experimentally into transition rules on
this discrete space \cite{m1}.   These discrete rules are merely tools for the
approximation of a continuous model, and other discretizations are possible.

  The agent (referred to as an ``ant") will be allowed to move from site to
site on a square lattice.    We allow each ant to take one step on the lattice
of
points (cells) at each time step.  As a result of discretizing the space, an
individual ant at each time step finds itself in one of these cells, and its
sensory input is the concentration of pheromone in its own cell
and each of the eight neighboring cells. In addition, each ant leaves
a constant amount of pheromone at the node in which it is located at every time
step. This pheromone decays (is reduced by a certain percentage) at each time
step. Toroidial boundary conditions are imposed on the lattice to remove, as
much as possible, any boundary effects.  The transition rates from cell $i$ to
cell $j$ are proportional to $W\propto\exp(-\beta (U[\sigma_j])$.  The
normalized transition probabilities on the lattice to go from cell
$k$ to cell
$i$ are then given by
\begin{equation}
P_{ik} =  {W(\sigma_i) w(\Delta_i)\over
\sum_{j/k}   W(\sigma_j) w(\Delta_j)}
\end{equation}
where the notation $\Sigma_j/k$ indicates the sum over all the cells $j$ which
are in the local neighborhood of $k$.  $\Delta_i$ measures the magnitude of the
difference in orientation (direction) to the previous direction the last time
the ant moved.  Since we are using a neighborhood composed of the cell and its
 eight neighbors on a square lattice, $\Delta_i$ can take only the
discrete values $0-4$, and it is sufficient to assign a number $w_i$ for each
of these changes of direction.  Here we used weights of (same direction)
$W_0=1$, and $w_1 = 1/2$, $w_2 = 1/8$, $w_3 = 1/12$ and $w_4 = 1/50$ (u-turn).
Once the parameters $\beta$, $\delta$ and $w_i$ are set  a large number of
ants can be placed on the lattice at random positions, the
movement of each ant can be determined randomly taken from the distribution
given
by $P_{ik}$.  We usually take the initial condition of the pheromone to be zero
at every point on the lattice.  Every time step each ant  leaves a quantity
$\eta$ (here $\eta = 0.05$) of pheromone  in each cell, and the total amount
of pheromone $\sigma_i$ in each cell is decreased at a rate
$\kappa$ (here $\kappa = 0.015$) at the end of each time step.  The evolution
of
the system is simulated numerically and the pattern of lattice sites that
contain ants is displayed in the figures.

\section{The Physiological Phase-Plot}

 We studied a wide range of scent laying and decay rates, as well as
different densities of ants.    This
 is the simplest  local, memoryless, homogeneous and isotropic  model which
leads to trail forming that we know of, and the formation of trails and
networks of ant traffic is not imposed by any special boundary conditions,
lattice topology, or additional behavioral rules.  The required behavioral
elements are stochastic, nonlinear response of an ant to the scent, and a
directional bias.  Furthermore the parameters of the system need to be tuned
somewhat   to the
appropriate region, and not any nonlinear rule of the above type will do.  If
the nonlinear response or the directional bias are removed no lines form, and
lines that are already formed do not persist.

Well defined trails form in the region above the phase transition line (ordered
phase), but near the transition from disorder. Further away from the
order-disorder line the clumping tendency overcomes the directional bias, and
no
lines form.

In every run observed within the part of the parameter space indicated in
Figure 4, a final system of trails emerged and survived as long as the
simulation
was run (Fig. 1). The self-organization of regular patterns of movement of the
ants occurs in two stages: an initial condensation phase, which occurs during
approximately the first 500 time steps, and a simplification phase. In the
first
stage a network of trails forms within a few hundred time steps. The trails at
first  typically form a network of many branches. All of the branches initially
appear stable, but after several hundred time steps, some abruptly disappear,
decreasing the length of the network.

In this second stage the network of ant traffic resembles qualitatively some
network type experiments done with real ants\cite{d3} and the network theory of
\cite{m1}, (and the basic discussion of the properties of these networks
given there holds). The ants can neglect a branch by chance, and if the
fluctuation in the density of the ants over the branch happens to be great
enough, the neglect builds on itself and the trail disappears.  As a result,
the
complexity of the network tends to decrease with time until a final, stable
network is reached.

The final line in all runs has been a loop (sometimes with
sub-loops). Most often, the trail exploits the periodic boundary
conditions by forming a more-or-less straight line that wraps around
the torus. This of course reflects the situation that a line ``free end"
is clearly unstable due to the directional bias.   That lines which feed back
on themselves invariably form is thus not surprising.   It is interesting to
note
here the qualitative similarity of these loops to  autocatalytic
sets of chemical reactions.\cite{k,f}  The instability of the free ends just
illustrates the sensitivity to boundary conditions when the information flows
in
an active way.  The separate issues of how these basic structures are
incorporated into a larger colony context will be taken up elsewhere, and we
believe they deserve further mathematical analysis.

The main conclusion to be drawn here is that simple osmotropotaxic scent
following of the very simple kind described above is no only sufficient to
allow for trail following behavior as shown in \cite{o2,o3}, but {\it
sufficient
to produce  evolution of complex pattern of organized flow of social insect
traffic all by itself}.

\section{Consolidation, Memory, and Bootstrapping}

Since the self-organizing properties of the swarm are instability driven, the
structures that form have some very interesting properties with respect to
large
perturbations.  We performed the following experiment:  the system was tuned to
a
region of the phase plane where lines form, and a stable network of traffic in
a
straight line was set up.  Then the noise level was increased so that  system
is
tuned below  the transition line ($\beta<\beta_c$).
 One observes that the ants fall away from their orderly patterns and
immediately
start executing random walks on the lattice.  As a result the pheromone
distribution starts to fluctuate and become more and more random.  If $\beta$
is then tuned back to its original value at some time later the line will
eventually reform with little or no change.  This occurs even if the
randomization is allowed to proceed to the point that the pheromonal field is
almost totally randomized.

Fig. 3 is shows a measure of how much of the original pattern in the
pheromonal field is left over time as the ants execute a random walk for a
certain time and then go back to nonlinear scent response. We define $\pi$ as
the
number of lattice sites where the concentration of pheromone is above the
average for the grid in both the original pattern and the current field, as a
percentage of the number of current sites with above average pheromone.
Thus $\pi$ is a measure of the overlap of the current pattern with the original
pattern.  The figure plots $\pi$ for different durations of the random walk,
and shows the stable patterns that re-form after the nonlinear scent response
is turned back on.
For times up to about twice the decay time $\tau =
1/\kappa$ even  small, virtually undetectable memory effects of the field can
be
amplified causing the patterns to reform without significant changes.  As shown
in Fig. 3, as much as 90\% of the pattern can be erased, and entirely
reconstituted later.

To further show the adaptability of the swarm in certain regions of the
parameter
space, the following experiment was performed. The system was tuned to a region
of $\beta$-$\delta$ parameter space where trails
never form, and the motion of the ants remains random. A weak
``bootstrapping" trail was then added to the grid, the pheromone density of
which was on the order of some of the larger random fluctuations in the field.
In a region slightly below the phase transition, this causes the swarm to
organize and amplify the trail, in spite of the fact that a network would not
have formed spontaneously. Fig. 4 plots a measure of pheromone concentration
$\upsilon$, which is the ratio of lattice sites with below-average pheromone
concentration to those with above-average concentration.  Three runs are shown;
one where no bootstrapping trail is added; another in which a bootstrapping
trail causes the swarm to shift into the trail, and lastly, an attempt at
bootstrapping which failed because the swarm was too far from the phase
transition line.  We emphasize that this behavior appears below, but near
enough to the phase transition line.

Memory reconsolidation and bootstrapping  are two seemingly conflicting
functional abilities which appear in the vicinity to the phase transition line.
Firstly, these organized patterns of behavior are really quite stable with
respect to large perturbation,s which might have an obvious usefulness to
operation in a changing and unpredictable environment.  Secondly, because the
patterns are due to the formation of an initially weak stable cooperative
structure, the swarm can act as an information amplifier, and  even a weak
external perturbation (such as the location of a food source by a single ant)
might lead to a significant response.  Thus swarms posses both a long memory
and the ability to learn.

\section{Conclusions}

As  shown in Fig. 2, lines of traffic form near the second-order phase
transition line.  In this region there is a cooperative effect
between the fluctuations and the ordering effects.   Much below the transition
line the fluctuations are too great, and no cooperative
structures form.   However, too far above the transition line the fluctuations
are suppressed, and order dominates overwhelmingly.  This means that the ants
are basically induced to ``turn around" with sufficient probability that   only
patches form.   The role of such  ``U-turns", has been
investigated experimentally and theoretically in \cite{ut}, where it was also
suggested that they might play an important role in the
self-organization of ant traffic.

The region in which the lines form represents a cooperative effect between the
fluctuations and the ordering behavior.  The formation of lines can
also be understood in terms of the theory in \cite{m1,m2}.  In the region
of the lattice where there are lines, the clumping effect is both unstable
in the transverse direction and stable in the lateral direction of the particle
motion, and it is this fact which gives rise to the formation of stable lines
of
traffic.  The result is a fairly robust region of line formation.  Motion in
this region can perhaps be likened to motion in the liquid crystal phase of
matter, or the reptation of polymer chains in polymer
melts, in which the order along one axis is frozen in, while along
another  motion can occur freely.

Since information (in the form of the orderly movement of
ants from one place to another) can be said to flow  only in the region of
stable
 line formation,
which lies above the disordered region  and below the very ordered region,
these results might mistakenly be thought similar to the ``meta-theory" of
``complexity at the edge of chaos" which asserts that complex behavior emerges
in the vicinity of a marginally stable state.\cite{bak1,pak,bak2,chris}
However
we point out that the system, as far as we know, does not ``self-organize" to
this region unless it may be said that on the evolutionary time scale the
biological organisms found such behavior adaptive.  In addition there is not an
``edge" but rather a large robust region where the most complex structures
form. Lastly we do not believe that the reason for this type of
behavior has anything to do with the very speculative ones which are sometimes
suggested.  It is quite natural that complex behaviors resulting from the
competition between stable and unstable modes should appear near a phase
transition line, since the first unstable modes  appear there.  Such behaviors
can be incorporated as biological functionality as shown in the few
examples discussed in this letter, but this functionality has nothing to
do with the notion of complexity ``at the edge of chaos" and the
hypothetical universal computational properties which may or may not exist
there.\cite{pak,chris}

ER would like to thank the the Santa Fe Institute and  the NSF REU program, and
the UGS program at LANL which supported parts of this research.

\begin{figure}[t]
%\vspace{2.8in}
%\hspace{.45in}\special{postscriptfile formation scaled 600}
\caption{\sf Trail formation.  A: shows the swarm at time 29, when the field is
still mostly random. B: at  $t=208$ C: at  $t=332$, the network has initially
formed, D: at $t=412$.  E: at $t=785$, where  some branches have disappeared.
F: at $t=1753$, shows the final network that emerges. }  \end{figure}

\begin{figure}[t]
%\vspace{1.5in}
%\hspace{0in}\special{postscriptfile figure.ps scaled 600}
\caption{\sf Physiological phase plot in $\delta$-$\gamma$ space for $\sigma_0
=1.367$. Figure show the distribution of particles after 5000 time steps
centered on the values of the physiological parameter for that run. }
\end{figure}

\begin{figure}
%\vspace{4.5in}
%\hspace{.55in}\special{postscriptfile consolidation.ps scaled 600}
\caption{\sf Memory erasure and reconsolidation. A: The original pattern.
B: $\beta$ is turned back up after 250 time steps; the trail re-forms
completely.  C: $\beta$ is turned back up after 265 steps; only a portion of
the
trail remains. D: $\beta$ is turned back up after 275 steps. E: $\beta$ is
turned back up after 300 steps. F: $\beta$ is turned back up after 375 steps. }
\end{figure}

\begin{figure}
%\vspace{4.5in}
%\hspace{1.2in}\special{postscriptfile bootstrapping.ps scaled 600}
\caption{\sf The ``bootstrapping" effect. The Plot
show the onset the effect of the introduction of a bootsrapping trail. Plot A
shows the random configuration of the ants before the trail is added. B: shows
the swarm in the short period of amplifying this weak trail. C: the final
state.
}
\end{figure}


\begin{thebibliography}{99}

\bibitem{Wilson} E. O.  Wilson, {\it The Insect Societies}, Cambridge:
Belknap Press, 1971; B. H\"olldobler, and  E. O. Wilson, {\it The Ants},
Cambridge:  Belknap (1990).

\bibitem{hang}  W. Hangartner, {\it Z. vergl. Physiol.} {\bf 57}, 103 (1967).

\bibitem{o2} V.  Calenbuhr and J.-L. Deneubourg,
{\it J. Theor. Biol.} {\bf 158}, 359 (1991).

\bibitem{o3} V.  Calenbuhr, L. Chr\'etien, J.-L. Deneubourg and C. Detrain,
{\it J. Theor. Biol.} {\bf 158}, 395 (1991).


\bibitem{d1} R. Beckers, J.-L. Deneubourg, S. Goss. and J. M. Pasteels, {\it
Insectes Soc.} {\bf 373}, 258 (1990).

\bibitem{d2} J.-L. Deneubourg, S. Aron, S. Goss and  J. M. Pasteels, {\it J.
Insect Behav.} {\it 32}, 159 (1990).

\bibitem{d3}  S. Goss, R. Beckers, J.-L. Deneubourg, S. Aron, and  J. M.
Pasteels,
In: {\it Behavioral Mechanisms of Food Selection, Nato ASI Series G20} (Hughes,
ed.) Berlin Heidelberg: Springer-Verlag (1990).

\bibitem{o1} V.  Calenbuhr and J.-L. Deneubourg, In:{\it Biological Motion} (W.
Alt, and G. Hoffmann, eds.) 453, Berlin: Springer-Verlag (1990).


\bibitem{m1} M. M. Millonas, {J. Theor. Biol.}  {\bf 159}, 529 (1992).

\bibitem{m2}  M. M. Millonas, In: {\it  Artificial Life III}(ed.  C. G.
Langton). Santa Fe Institute  Studies in the Sciences of Complexity, Proc. Vol
XVII. Reading,  Massachussetts: Addison-Wesley, (1994).

\bibitem{m3} M. M. Millonas, In:  {\it Pattern Formation in Physical and
Biological Sciences} (ed. P. Claudis).  Santa Fe Institute Studies in the
Sciences of Complexity, Reading Massachussets: Addison-Wesley (1994).

\bibitem{ww2} E. O.  Wilson, {\it Animal Behavior} {\bf 10}, 134-164 (1962).


\bibitem{p1}  R. P. Evershed, E. D. Morgan and  M. C. Cammaerts, {\it Insect
Biochem.} {\bf 12}, 383 (1981).

\bibitem{p2} C. Detrain, J. M. Pasteels, and J.-L. Deneubourg, {\it
Actes coll. Insectes Sociaux} {\bf 4}, 87 (1988).

\bibitem{p3} S. G\'erardy and J. C. Verhaeghe, {\it
Actes coll. Insectes Sociaux} {\bf 4}, 235 (1988).

\bibitem{other}   \bibitem{} L. Lam and R.D. Pochy, Compt. Phys. 7, 534 (1993);
Pochy, D.R. Kayser, L.K. Aberle and L. Lam, Physica D 66, 166 (1993);
F. Schweitzer and L. Schmansky-Geier, {\it Physica A} {\bf 206}, 359 (1994).

\bibitem{hakken}  H. Haken, Synergetics, Berlin: Springer-Verlag (1983).

\bibitem{k}  S. A. Kauffman, {\it J. Theor. Biol.} {\bf 119 }, 1 (1986).

\bibitem{f}  J. D. Farmer, S. A. Kauffman and N. H. Packard, {\it Physica D}
{\bf 22}, 50 (1986).




\bibitem{ut} R. Beckers, J.-L. Deneubourg and S. Goss,
{\it J. Theor. Biol.} {\bf 159 }, 397 (1992).

\bibitem{bak1}  P. Bak, C. Tang and K. Weisenfield, {\it Phys. Rev. A} {\bf
38},
364 (1988).

\bibitem{pak}  N. H. Packard, In: {\it Complexity in Biological Modeling} (eds.
S. Kelso and M. Shlesinger) (1988).

\bibitem{bak2}  P. Bak, K. Chen and M. Creutz, {\it Nature}  {\bf 342}, 780
(1989).

\bibitem{chris}  C. G. Langton, {\it Computation at the edge of chaos: Phase
transitions and emergent computation}, Ph. D. Thesis,  University of Michigan,
1991.







\end{thebibliography}
\end{document}